# Design of a molecular Field Effect Transistor (mFET)

Ralph C. Merkle, Robert A. Freitas Jr. and Damian G. Allis
Institute for Molecular Manufacturing

## Abstract

Field Effect Transistors (FETs) are ubiquitous in electronics. As we scale FETs to ever smaller sizes, it becomes natural to ask how small a practical FET might be. We propose and analyze an atomically precise molecular FET (herein referred to as an "mFET") with 7,694 atoms made only of hydrogen and carbon atoms. It uses metallic (4,4) carbon nanotubes as the conductive leads, a linear segment of Lonsdaleite (hexagonal diamond) as the channel, Lonsdaleite as the insulating layer between the channel and the gate, and a (20,20) metallic carbon nanotube as the surrounding gate. The (4,4) nanotube leads are bonded to the channel using a mix of 5- and 6-membered rings, and to the gate using 5-, 6- and 7-membered rings. Issues of component design assessment and optimization using quantum chemical methods are discussed throughout. A 10 watt sugar-cube-sized computer made with $10^{18}$ such mFETs could deliver ~$10^{25}$ switching operations per second.

# Introduction

Field Effect Transistors (FETs) are central to the computer industry and for most electronic devices now in use. They are used to detect and amplify electronic signals and to perform basic logic operations on electronic signals representing the digital ones and zeros of binary logic. FETs have only a few major components [1-5]. These are the source (S), the drain (D), the channel (C) that connects the source and the drain, the gate (G), and the insulator (I) between the gate and the channel.

A conventional planar FET and a FinFET are shown in Figure 1.

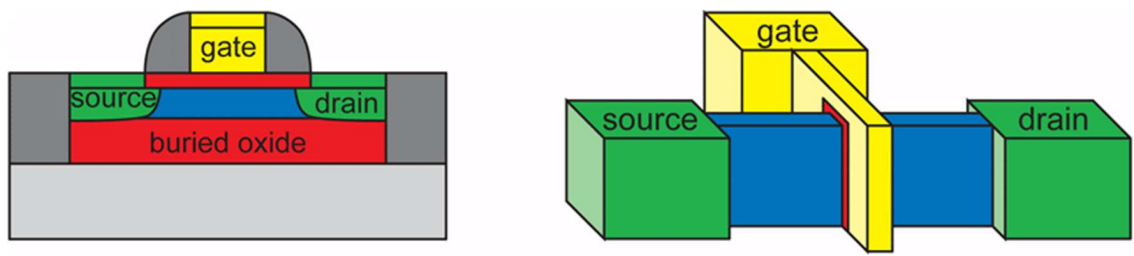

*Figure 1. Left: a conventional FET. Right: a FinFET. Image from [5].*

In operational macroscale FETs, charge carriers are in high concentration in the source and drain, and continue to conduct current even in the presence of a significant bias voltage on the gate. Conversely, charge carriers are in much lower concentration in the channel, usually because of a low concentration of dopant atoms in that region. As a consequence, the presence of a bias voltage on the gate (a source-gate voltage difference) can significantly increase or decrease the concentration of charge carriers in the channel.

If the number of charge carriers in the channel is reduced, the channel will conduct less current. If the number of charge carriers in the channel is increased, the channel will conduct more current.

If there are too many charge carriers in the channel, their sheer numbers will swamp the bias voltage applied to the gate and the channel will continue to conduct current. We can limit the number of charge carriers if (a) the channel region is mostly unoccupied (it is mostly vacuum) or if (b) almost all the electrons in the channel region occupy orbitals that are spatially confined, and those electrons have insufficient energy to reach the next available (open) orbital, making the entire channel unable to conduct current. In the former case, we have a "vacuum tube" of some sort, referring to a device that uses electrons in a vacuum whose current flow between a source (usually called a "cathode") and a drain (usually called a "plate") is modulated by an electric field (usually applied by a "grid") to amplify signals. In the latter case, we refer to the spatially confined electrons in the channel as "occupying the valence band" and the unconfined electrons as "occupying the conduction band," while the device as a whole is usually referred to as some sort of FET or, more generally, a semiconductor device.

While the planar FET depicted in Figure 1 is the most widely known design, the ultimate geometry is likely to be a cylindrical "gate-all-around" (GAA) design shown in Figure 2. As described in the first report demonstrating a fabricated GAA FET by Chen et al. of IBM, Harvard, and Purdue [4]:

> "Single-wall carbon nanotubes (CNs) [CNTs] are considered to be one of the most promising candidates for post-CMOS applications, mainly owing to their smallness and ballistic transport properties. The ultrathin body of CNs (of the order of a few nanometers) allows for aggressive channel length scaling while maintaining excellent gate control. In general, a gate-all-around (GAA) structure is expected to be the ideal geometry that maximizes electrostatic gate control in FETs. Combining the ultrathin body of a CN with a GAA device geometry is a natural choice for ultimate device design".

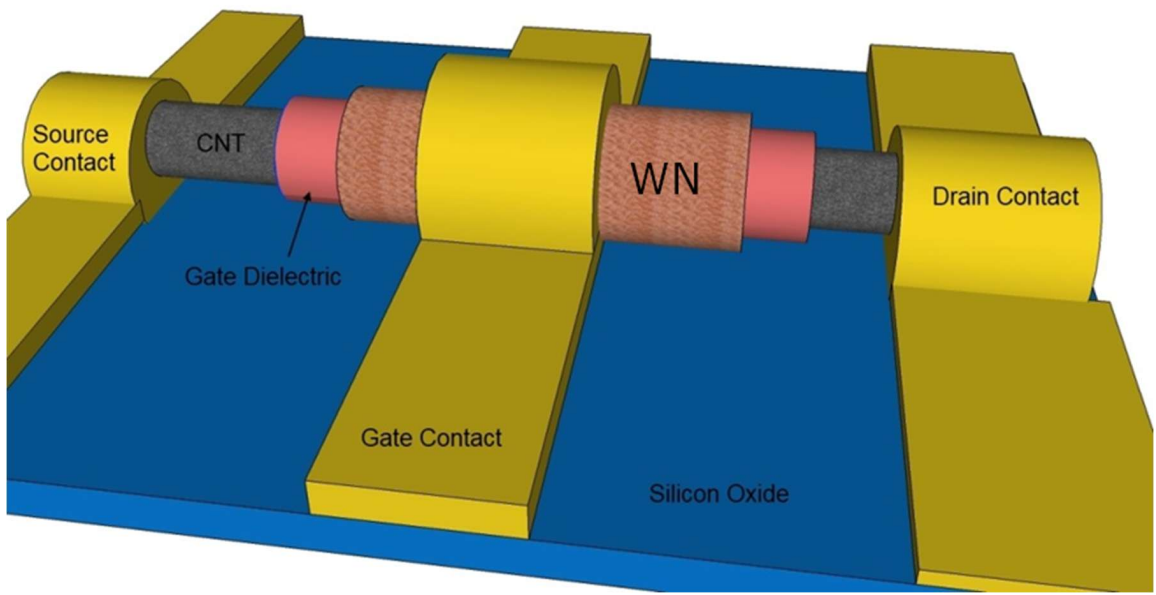

*Figure 2. Illustration of a Gate-All-Around (GAA) FET, the ultimate FET geometry (illustration from [4]). WN = Tungsten Nitride (metal gate); CNT = a CNT functionalized with $NO_2$ and wrapped in an $Al_2O_3$ dielectric using atomic layer deposition. The gap between the source and drain in the first fabricated GAA FET is approx. 250 nm, of which the gate contact accounts for approximately 100 nm.*

The basic concept of the carbon nanotube (CNT) GAA FET is simple. The channel consists of a semiconducting nanotube running straight through the center of the device. The gate, as the name implies, wraps around the device. The insulating layer wraps around the channel and separates the gate from the channel. When a voltage is applied to the gate, the voltage throughout the region inside the cylindrical gate (and reasonably far from the ends of the cylindrical gate) equilibrates to the voltage on the gate. That is, the GAA FET applies the same voltage to the entire channel region that is applied to the gate, rather than applying some constantly falling fraction of the gate voltage to the channel region – which is what occurs in planar FETs.

As the authors of [4] say, "a gate-all-around (GAA) structure is expected to be the ideal geometry that maximizes electrostatic gate control in FETs." At the same time, we want the smallest possible FET for the simple reason that smaller FETs have less parasitic capacitance and hence a higher frequency of operation. The smallest size can be achieved when the channel is as thin and short as possible.

## Channel

While difficult to say what the thinnest possible channel might be, low diameter nanotubes are very thin and have already been used successfully as the channel in FETs. Generally, CNTs can be conveniently divided into metallic or non-metallic categories based on a simple structural parameter. Given the standard indices (*n*,*m*) used to describe the chirality of a CNT, we compute (*n* - *m* mod 3). If this is 0, the CNT is metallic. If not, the CNT is not metallic and has some bandgap. An approximate formula for the minimum bandgap of a non-metallic CNT is $E_{gap} \sim 0.8eV/D[nm]$ [6,7]. For the (4,0) CNT with D ~ 0.3 nm, this formula gives $E_{gap} \sim 2.7$ eV. Unfortunately, while the (4,0) and other very small diameter CNTs are formally semiconducting, the approximate formula given above breaks down due to the loss of full electronic delocalization that comes from high strain at the aromatic C=C bonds in the highly curved tubes, and more accurate *ab initio* calculations predict a negligible bandgap. In the interest of greatly reduced device size and considering the complexities of the electronic structure of the smallest possible chirality options, the use of CNTs for the channel might not be the best choice.

Diamond, on the other hand, has a substantial bandgap (~5.5 eV) and a long, thin piece of diamond should make for an excellent channel. Unfortunately, diamond has negative electron affinity, which means if we use electrons as the charge carriers in a diamond channel (that is, we have an N-type FET) most electrons in the channel will exit from the sides of the channel rather than through the source or drain. This is unlikely to result in acceptable FET function. On the other hand, holes (the absence of electrons in the valence band) can be used as charge carriers and are unlikely to leave the confines of the channel. They will instead travel along its length, entering and exiting through the source or drain. Accordingly, a P-type FET using a very narrow diamond channel should be realizable.

Following this observation, we computed the properties of a candidate channel material - a linear segment of hexagonal diamond, also known as Lonsdaleite - as presented for the channel in Figure 3 and as illustrated from density functional theory (DFT) calculations in Figure 4. Its trend in bandgap, as computed with Gaussian16 [8] at the B3LYP/6-311G(2d,p)//B3LYP/6-31G(d,p) level of theory [9-12] under both molecular (to show the change in highest occupied and lowest unoccupied molecular orbital (HOMO-LUMO) gap with increasing length) and periodic boundary conditions (to better approximate the properties of a much longer version of this channel) is shown in Table 1. Notably, the calculated bandgap is, at 6.4 eV (calculated), significantly larger than the ~5.5 eV bandgap of bulk diamond. The negative electron affinity is computed to be 0.67 eV. This means the conduction band minimum of the candidate channel lies 0.67 eV above the vacuum level. The work function, or ionization energy, for removing an electron from the valence band of the candidate channel as calculated is 5.8 eV. As the work function of CNTs has been experimentally determined to be ~5 eV (though this value varies depending on chirality), this means that an applied positive source-gate bias (that is, the source has a positive voltage with respect to the gate) of slightly more than 0.8 V should generate holes in the source end of the channel. Predicted work functions for (4,4) nanotubes of ~ 4.5 eV have been computed [13], which would imply a bias of 1.3 V. Increasing this bias should increase the rate at which holes are injected into the channel by the source. While we eschew elements other than carbon and hydrogen in this initial design, P-doping of the channel near its junctions with the (4,4) nanotubes would be a more traditional method of providing low resistance ohmic contacts. A few substitutional boron atoms might also suffice.

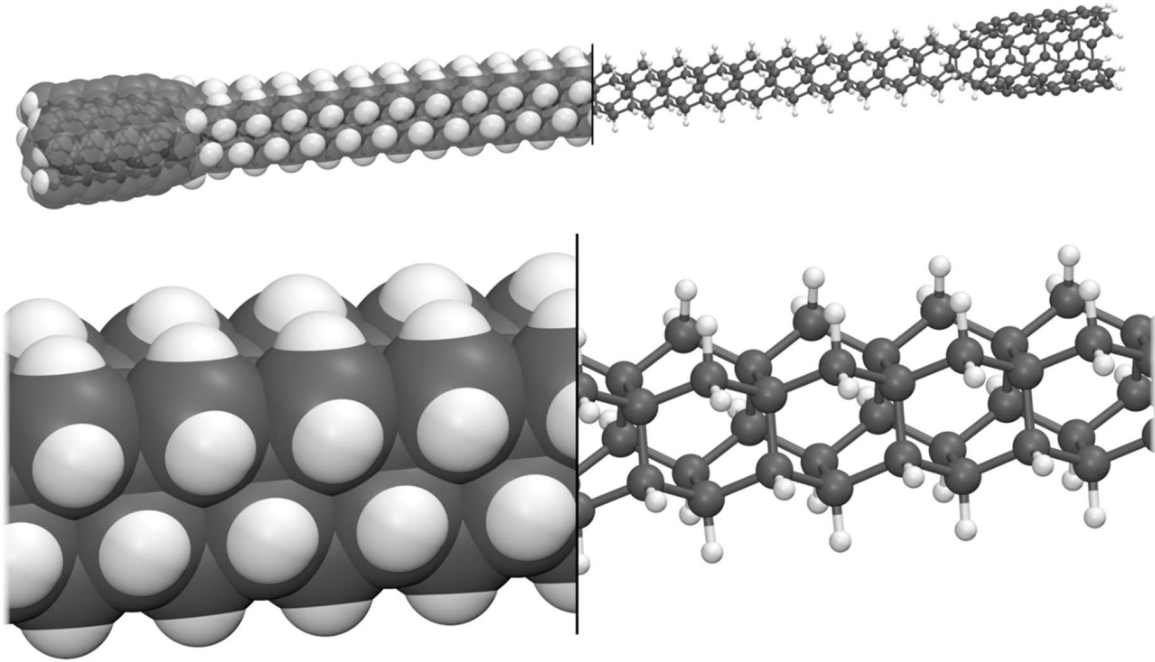

*Figure 3. Top: Lonsdaleite channel with attached (4,4) CNT leads (502 atoms). Bottom: close up of the channel in van der Waals (left) and ball-and-stick (right) representations.*

Table 1. Indirect electronic band gap calculations of the Lonsdaleite channel at the B3LYP/6-311G(2d,p)//B3LYP/6-31G(d,p) level of theory. The calculated 6.4 eV band gap is the estimate for the channel at significant ( > 10 unit → "infinite") lengths. PBC: Periodic Boundary Conditions.

| Lonsdaleite Repeat Units | Band Gap (eV) |
|---|---|
| 1 (molecular) | 7.961 |
| 2 (molecular) | 7.393 |
| 5 (molecular) | 6.825 |
| 10 (molecular) | 6.571 |
| PBC (Unit Cell, 1x) | 6.521 |
| 20 (molecular) | 6.462 |
| PBC Supercell (4x) | 6.428 |

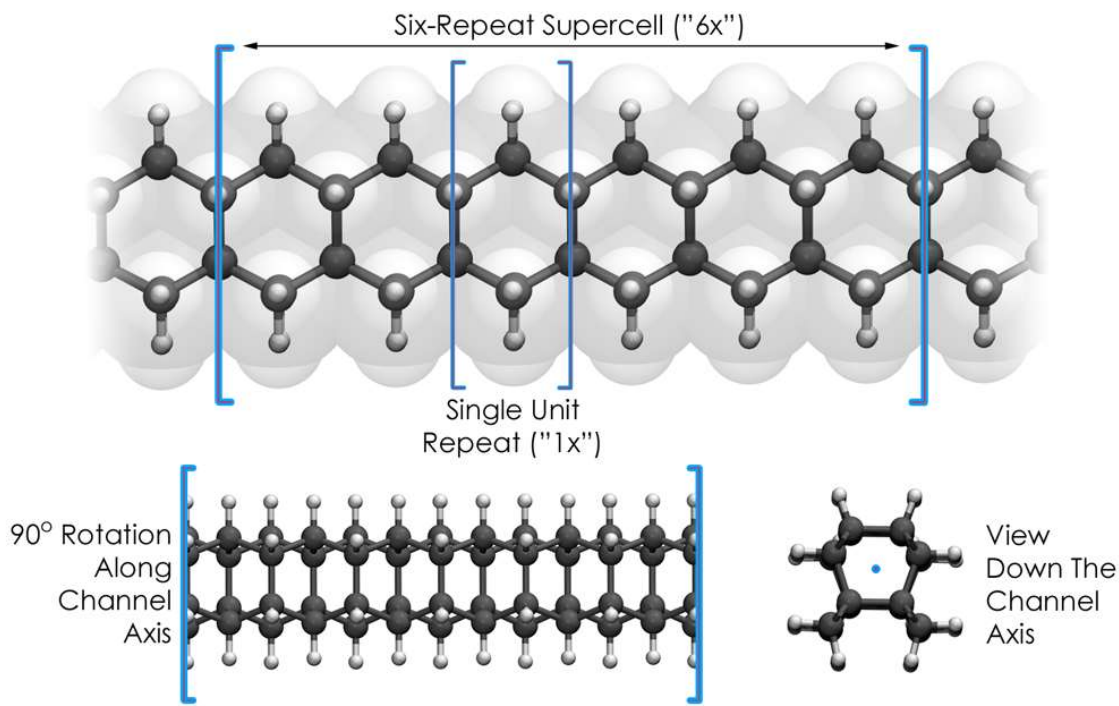

Figure 4. Three views of the proposed Lonsdaleite channel, including identification of the single repeat unit and representative six-unit supercell. The blue dot at lower-right is the channel axis.

In other words, a molecular P-type FET in which metallic nanotubes are connected to a one dimensional segment of hexagonal diamond used as a channel should function quite well when used in a regime in which a source-gate bias voltage of somewhat more than 1 V causes current flow, and a source-gate bias voltage of 0 V or less prevents current flow.

## Insulator

We can most conveniently use a form of diamond as an insulator. Here, we will use hexagonal diamond, or Lonsdaleite, as it is easier to sculpt this form with its hexagonal symmetry into a cylinder than it is the cubic diamond lattice – all while still retaining a simple atomically precise crystalline structure (Figure 5). In addition, the dielectric strength of CVD diamond is 1 V/nm [14]. Femto Science claims an experimental demonstration of a dielectric strength of 3 V/nm (30 MV/cm) [15]. The distance from the gate, a (20,20) nanotube, to a representative inner channel structure, a (4,0) nanotube, is 1.2 nm (from nucleus to nucleus). Whether this is the proper "distance" metric to use is an interesting question. The van der Waals distance between two sheets of graphite is ~0.34 nm, and we might reasonably argue that this distance should be subtracted from the internuclear distance we just computed to get a better approximation to the "gap" through which electrons would have to jump, and therefore the thickness of the insulating layer of diamond involved. This "adjusted" distance would be 0.86 nm and would de-rate the maximum breakdown voltage of our device from 1.2 V to 0.86 V, assuming that the insulator was able to withstand an electric field of 1 V/nm (the approximate breakdown field strength of diamond). In any event, the likely breakdown voltage of this mFET will be in the vicinity of 1 V and possibly (for reasons discussed below) higher. This should be sufficient to allow reasonably reliable operation at room temperature, although if we want to reduce the source-drain leakage current even further (which might be desirable if we wish to adopt extremely low leakage currents to minimize energy dissipation) we might wish to double the voltage to ~2 V. This would double the linear dimensions of the device and multiply the atom count by approximately a factor of eight if we made no attempt to adopt a more sophisticated design. Methods of reducing the atom count of the device are discussed below. These would improve device performance both for the device as analyzed and for any larger device if operation at a higher voltage were desired.

With respect to the proposed design, and as a demonstration of the types of quantum chemical analyses that can be applied to systems of this size in the optimization of structural parameters (such as the quality-of-fit for interacting components) for the combinations of any similar parts that can be considered, reduced model channel-insulator combinations were assessed by density functional tight-binding (DFTB) methods [16].

DFTB+ [17] calculations using the PBC-0-3 parameter set [18] both with and without the Grimme D4 dispersion correction [19] reveal the importance of dispersion corrections in obtaining correct interaction energies (Table 2). The model structures used for the channel-insulator assessment (Figure 6) reveal a linear increase in favorable interactions with the PBC-0-3/D4 combination (approximately -0.75 eV per repeat unit), providing insights into quality-of-fit for these two components at a fraction of the cost of a full DFT calculation (of what would need to be even further-simplified structures to obtain these predictions in a reasonable timeframe). The calculations show that the inclusion of dispersion (van der Waals forces), a type of weak and long-range interaction that most density functionals do not account for and that is most often introduced into calculations through efficient and effective empirical corrections, makes the channel-insulator interaction energetically favorable (negative energy), indicating a good fit between the two components. This predicted trend is an indicator that this channel/insulator combination is likely to be an excellent fit for our purposes.

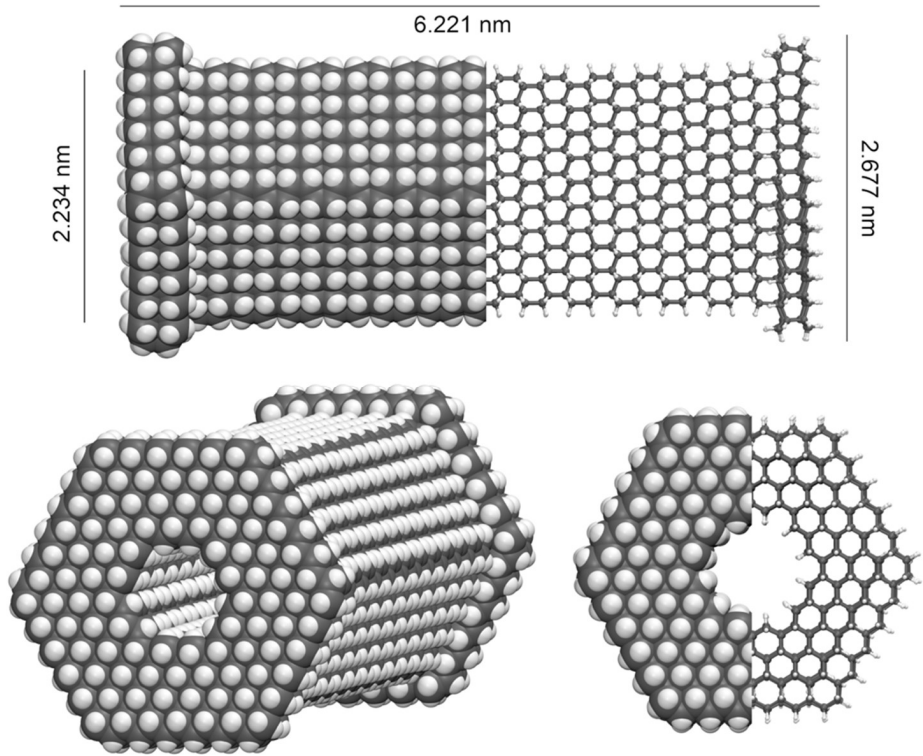

*Figure 5. Three views of the mFET Lonsdaleite insulator (5418 atoms).*

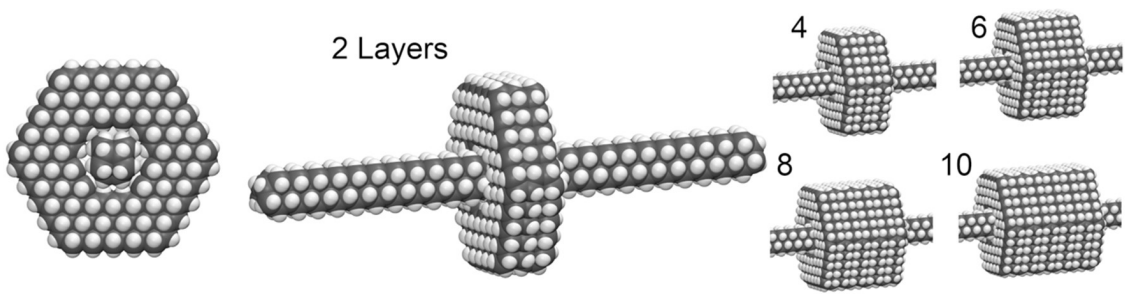

*Figure 6. Structures used in the assessment of the channel-insulator quality-of-fit.*

Table 2. Channel-Insulator electronic interaction energies with the PBC-0-3 parameter set both with and without D4 dispersion corrections. The fit is found to be net-stabilizing at this level of theory upon inclusion of the dispersion correction in a nearly linear manner per additional insulator repeat unit (approx. -0.75 eV). By convention, negative energies indicate favorable (stabilizing) interactions, while positive energies indicate repulsive (destabilizing) interactions.

| *Insulator Unit* | *PBC-0-3 Energy (eV)* | *Energy Per Repeat Unit (eV)* | *PBC-0-3 + D4 Energy (eV)* | *Dispersion-Corrected (D4) Energy Per Repeat Unit (eV)* |
|---|---|---|---|---|
| *2-Layer* | 1.2601 | 0.6300 | -1.8363 | -0.9182 |
| *4-Layer* | 1.7550 | 0.4388 | -3.1285 | -0.7821 |
| *6-Layer* | 1.9485 | 0.3248 | -4.6998 | -0.7833 |
| *8-Layer* | 2.3873 | 0.2984 | -6.0269 | -0.7534 |
| *10-Layer* | 3.2576 | 0.3258 | -6.9308 | -0.6931 |

For comparison, a significantly reduced channel-insulator geometry was constructed under symmetry constraints ($C_{2v}$) to allow for a direct DFT-to-DFTB comparison of the interaction energies for the larger system in Figure 6. Interaction energies for this model complex (Figure 7) were calculated at the B3LYP/6-311G(2d,p)//B3LYP/3-21G* and CAM-B3LYP/6-311G(2d,p)//B3LYP/3-21G* levels of theory both with and without the D3 version of Grimme dispersion with Becke-Johnson damping (GD3(BJ), [20]) and with the PBC-0-3 DFTB parameter set both with and without the D4 dispersion correction. These interaction energies are summarized in Table 3 and reveal that, for a very modest computational cost compared to full DFT calculations, DFTB provides for very reasonable estimates of electronic interaction energies, including demonstrating the importance of adding the dispersion corrections for obtaining the correct direction in favorable interaction energy for these small diamondoid model systems.

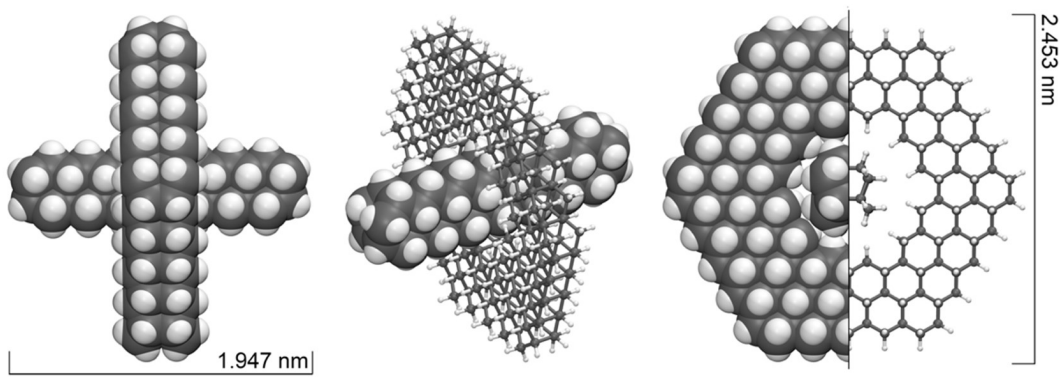

*Figure 7. The model channel-insulator structure produced for assessing the performance of the DFTB PBC-0-3 parameter set against full DFT calculations (see text). The overall structure and components were optimized under $C_{2v}$ symmetry constraints.*

Table 3. Electronic interaction energies (in eV) between the model channel and insulator shown in Figure 7. "B3LYP" = B3LYP/6-311G(2d,p)//B3LYP/3-21G*; "CAM-B3LYP" = CAM-B3LYP/6-311G(2d,p)//B3LYP/3-21G*. By convention, negative energies are net-stabilizing. On the same computer, the CAM-B3LYP-GD3(BJ) single energy calculation of the model complex took 115 hours. The entire PBC-0-3/D4 optimization required 1.5 hours.

|              | *B3LYP* | *CAM-B3LYP* |       | *PBC-0-3* |
|---          |---      |---          |---    |---        |
| *Uncorrected* | 1.4071 | 0.9369     |       | 0.2475    |
| *w/GD3(BJ)*  | -1.7902 | -1.6327    | *w/D4* | -2.0028   |

It is also worth noting that a device that is both this small and this perfect will not suffer a "breakdown" in the manner that more macroscale devices will break down. A macroscale device will inherently have multiple defects due to today's manufacturing methods and the inability to fabricate with atomic precision. Stray electrons will be accelerated over long distances, creating cascades of energetic electrons capable of damaging the lattice. This mFET and other atomically precise candidate devices will have no "stray" electrons to initiate a cascade of accelerated electrons, and a near-complete understanding of the electronic structure and excited states of all of the components can be obtained through quantum chemical analysis to address the behavior of electrons at the surfaces and within the carbon lattices themselves as part of the characterization and subsequent optimization of the system behavior. The electric field will continue to build until some electron flows through the structure. Even after this happens it is not at all clear that single electrons will accelerate into other electrons and cause a cascade. As the total distance travelled will be only a few nanometers, single electrons will only gain a few eV, not enough to dislodge other electrons. They would likely flow through without doing significant damage. The nature of a "breakdown" at this scale requires further investigation.

## Gate

The gate need only conduct charge around the mFET. To this end, we simply wrap the insulating layer of Lonsdaleite in a (20,20) or other CNT of appropriate radius and chirality (Figure 8). The (*n,n*) nanotubes are quite conductive (being metallic), and can be easily selected to be large enough to surround the insulator. They will not be called upon to conduct much current and require only a single layer of carbon atoms. In the proposed design, the (20,20) nanotube gate has been modified by bonding a (4,4) metallic nanotube to it. The (4,4) nanotube acts as a metallic conductive wire [6,7] that connects the gate to the external circuit. The electronic structure of this junction (specifically, the quality of orbital overlap between the (20,20) and (4,4) regions through which electrons will pass) or similarly adequate designs is under investigation.

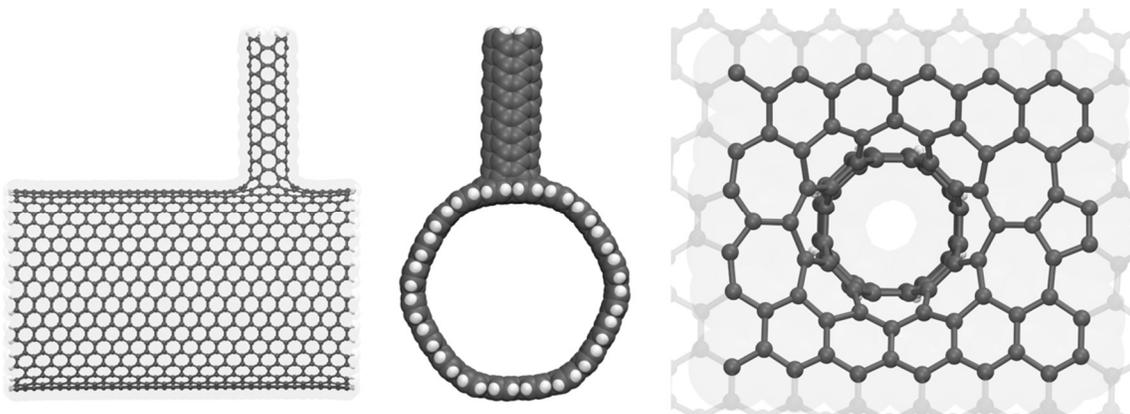

*Figure 8. Two views of the (20,20) CNT gate (1774 atoms) from the optimized mFET. At right, the connectivity of the (4,4) CNT to the (20,20) CNT.*

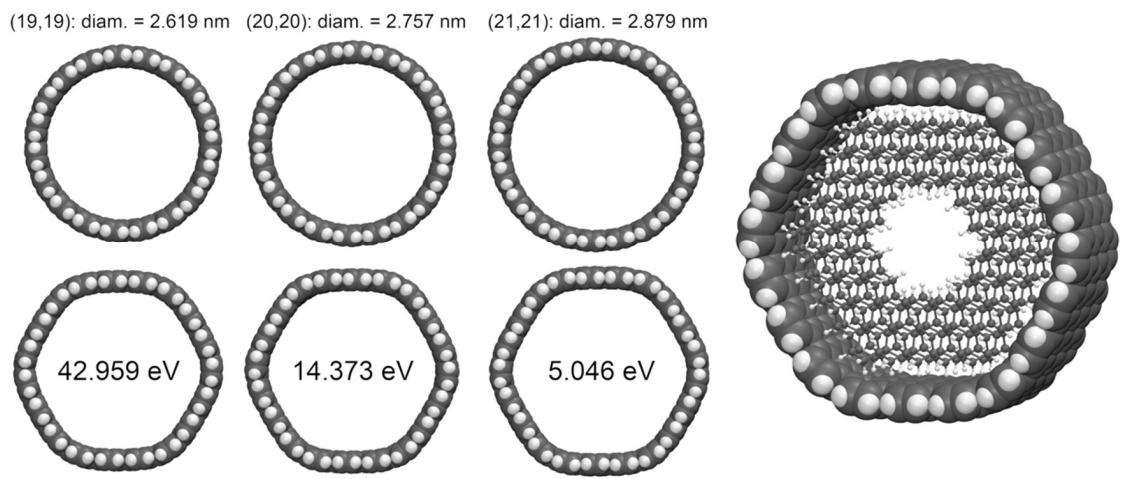

*Figure 9. The (19,19) (left), (20,20) (middle), and (21,21) (right) CNTs in isolation (top) and optimized with the Lonsdaleite insulator segment in place (bottom), resulting in hexagonal deformations. Values (in eV) within the hexagonally deformed CNTs are the energy differences between these and the optimized (cylindrical) CNTs for each chirality (representing the deformation of the CNTs due to interactions with the insulator) with the PBC-0-3/D4 parameter combination. At right, an example (20,20) CNT with the insulator repeat unit in place as used for the DFTB fit assessment for all three CNT/Insulator combinations.*

Again, DFTB analyses can be performed on the insulator/gate combination to provide insights into key structural parameters (Figure 9). Optimization of reduced model insulator/gate components with the PBC-0-3 parameter set and D4 dispersion correction indicate that the Insulator/(19,19) combination has a net-destabilizing energy of 24.2 eV – a large energy distributed largely through deformation of the CNT into a hexagonal configuration around the insulator. The combination of insulator and (20,20) CNT is net-stabilizing by 2.245 eV (a small amount over a large system). The (21,21) CNT is net-stabilizing by 8.722 eV. Larger sizes could be modeled, but the clear prediction is that no efforts to further tailor the CNT chirality or conductivity need be considered beyond the simple unit increase in (*n,n*) to achieve fits that are potentially destabilizing (which may result in shifts of molecular orbitals relevant to mFET operation) or slightly-to-somewhat stabilizing (which may have little-to-no impact but add to the structural integrity of the final design).

## Source/Drain Connectivity

The Lonsdaleite channel is connected at both ends to two (4,4) metallic nanotubes that act as source and drain leads. The (4,4) nanotube is stable in ordinary room temperature environments and does not require special handling. It can therefore be used to carry electrical signals between components and to connect components to external (possibly macroscopic) circuits. A molecular segment of the (4,4)-to-Lonsdaleite coupling is shown in Figure 10. The density-of-states (DoS) of this molecule, indicating a HOMO-LUMO gap (band gap approximation) of 2.19 eV at the B3LYP/6-311G(2d,p)//B3LYP/6-31G(d,p) level of theory and 3.98 eV at the CAM-B3LYP/6-311G(2d,p)//B3LYP/6-31G(d,p) [21] level of theory, is shown in Figure 11. This orbital energy variation among density functionals (and, here, HOMO-LUMO gap) is not unexpected – the two calculations are provided to (a) reveal that a range of predicted energies exists that additional computational studies can better address, (b) that this component is amenable to high-level computational studies to facilitate this experimental tuning, and (c) that, despite the predicted range of these two (of many available) density functionals (as well as available basis sets), the CNT frontier orbitals dominate the DoS in the energy range of interest for this system.

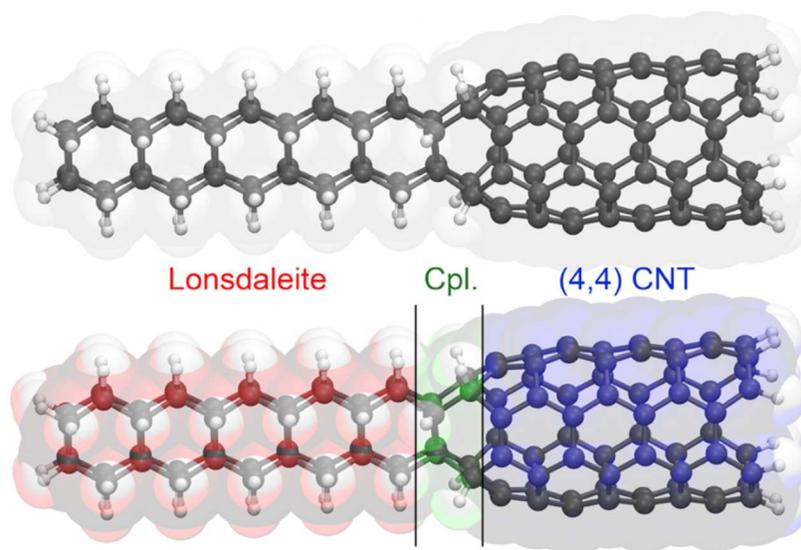

*Figure 10. The (4,4)-to-Lonsdaleite coupling (Cpl.) of the channel to its edge CNTs. The color scheme for the components is as also shown for the (p)DoS plots in Figure 11.*

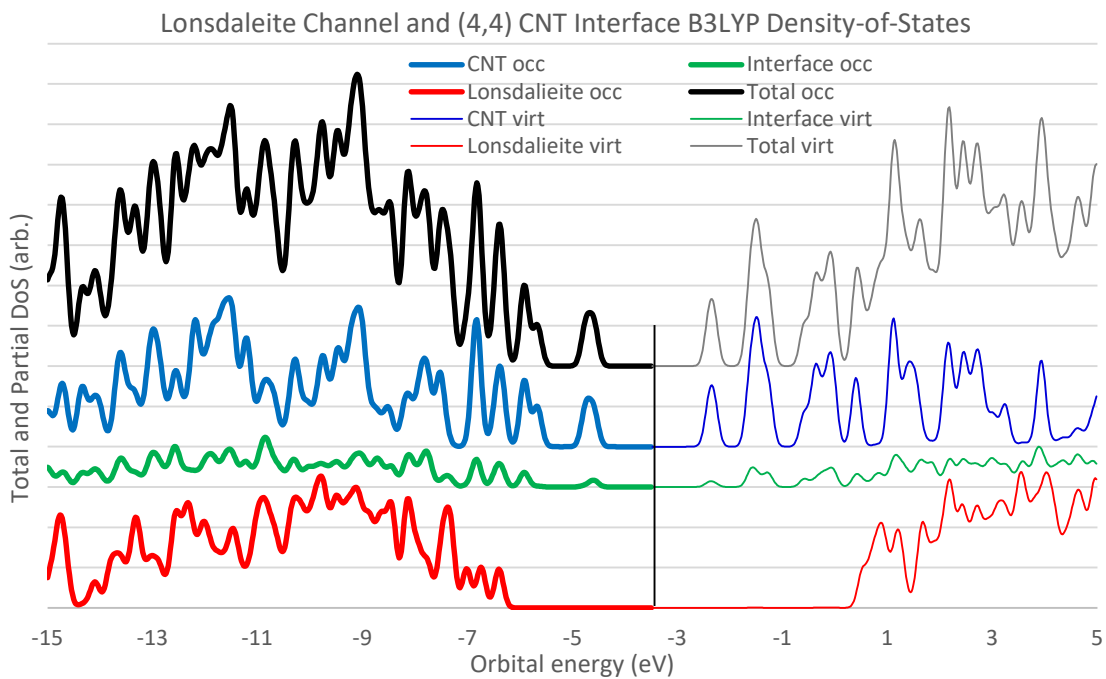

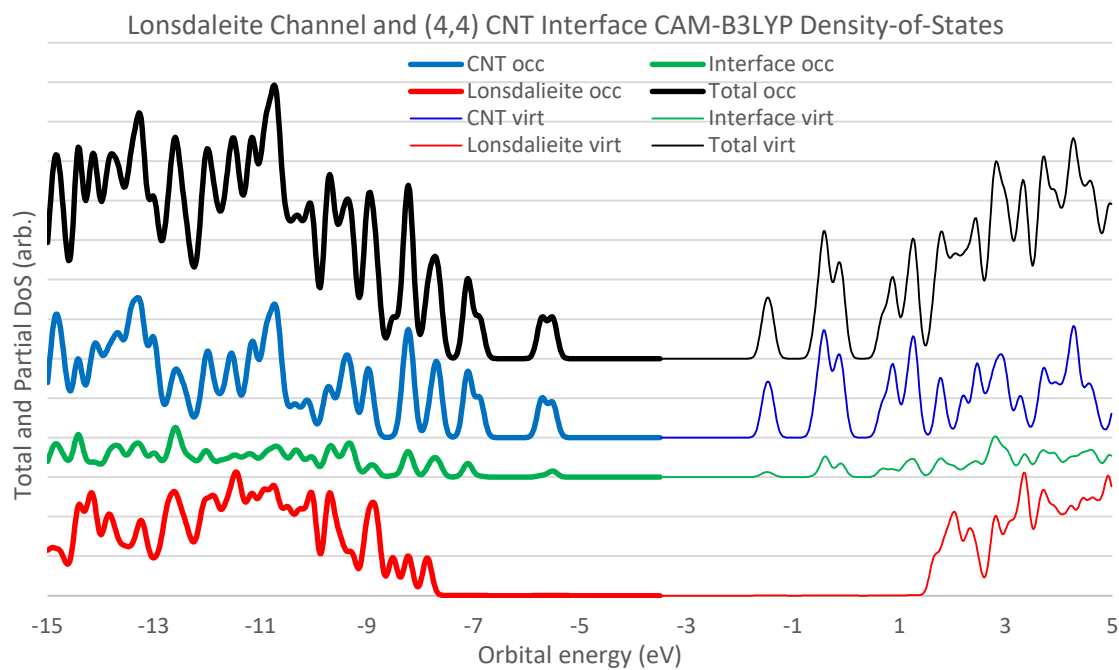

*Figure 11. Total and partial occupied (occ) and virtual (virt) density-of-states (DoS) for the Lonsdaleite channel, (4,4) CNT lead, and interfacing (coupling) carbon and hydrogen atoms between both subunits (see Figure 10). CNT orbitals are found to dominate around the HOMO-LUMO gap (HOMO: B3LYP = -4.579 eV; CAM-B3LYP = -5.501 eV; LUMO: B3LYP = -2.386 eV; CAM-B3LYP = -1.518 eV). Calculations performed at the B3LYP/6-311G(2d,p)//B3LYP/6-31G(d,p) and CAM-B3LYP/6-311G(2d,p)//B3LYP/6-31G(d,p) levels of theory. Plots are offset along the ordinates for clarity.*

The same chirality (4,4) nanotube is also connected to the gate, connecting the gate to (possibly macroscopic) external circuits. In the case of the source and drain, the use of (4,4) nanotubes is due to the symmetry and connectivity of the Lonsdaleite channel itself, whose ends are ideally suited for covalent bonding to the four available carbons at the proposed tapered end of this chirality. For the gate, any small metallic nanotube that can be covalently hybridized with the hexagonal framework of the (20,20) or other metallic nanotube to produce a continuous electron-delocalized framework should suffice.

## Assembled molecular Field Effect Transistor

The assembled mFET is shown in Figure 12, optimized with the ND-1 force field available within NanoEngineer-1 [22].

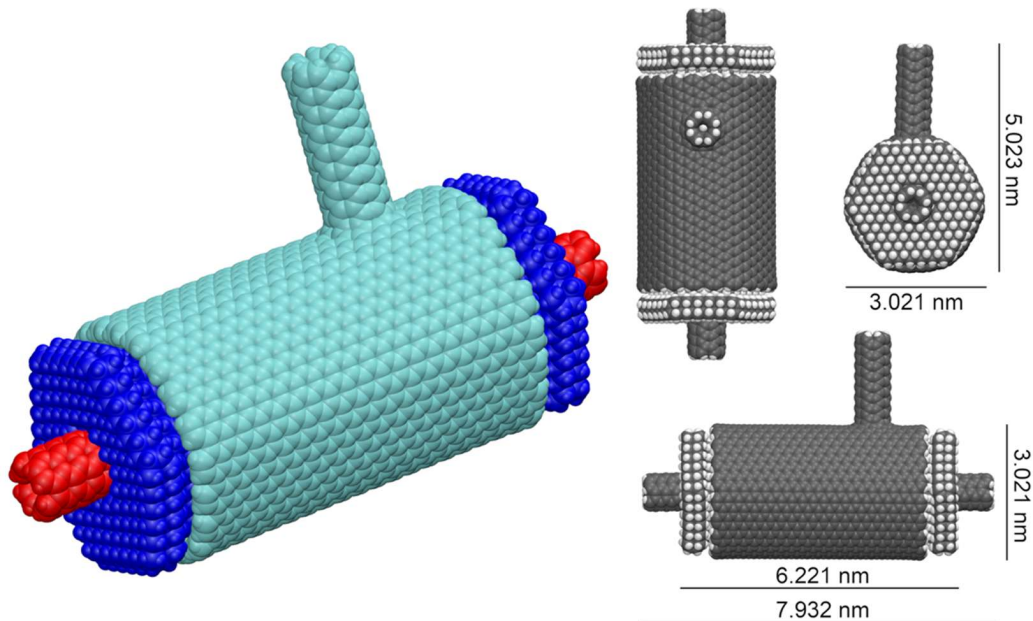

*Figure 12. Multiple views of the proposed mFET (7694 atoms), with colors at left used to distinguish all components in the design. The total volume of the mFET is 46.52 nm$^3$ based on van der Waals radii.*

## Parasitic Capacitance

This molecular FET is quite small and, consequently, has low capacitance. Diamond has a dielectric constant of ~5.7 [2]. The capacitance of a cylindrical capacitor is calculated as:

$$C = 2 \pi \kappa \varepsilon_0 L / \ln (b/a)$$

Where:
$\varepsilon_0$ is the permittivity constant, 8.85 x $10^{-12}$ F/m
$\kappa$ is the dielectric constant for the material between the plates
**L** is the length of the cylinder
**b** is the outer radius
**a** is the inner radius

$$b = 1.4 \text{ nm} - 0.17 \text{ nm}$$
$$a = 0.165 \text{ nm} + 0.17 \text{ nm}$$
$$\ln(b/a) = 1.29$$
$$L = 4.7 \text{ nm}$$

Which gives C ~1.2 x $10^{-18}$ F.

While this is approximate (this idealized formula for capacitance will have some significant error at this small a length scale) it gives some idea of the parasitic capacitance of our mFET. While more accurate calculations would give a more precise answer, the basic conclusion at this point is that the parasitic capacitance is, as expected, quite small.

It is worth noting that the dielectric constant of diamond is 5.7, but the dielectric constant of vacuum is 1. By replacing some of the diamond with vacuum, we could reduce the dielectric constant significantly. This can, for example, be accomplished by drilling out the block of Lonsdaleite used as an insulator. Candidate insulator structures are shown in Figure 13.

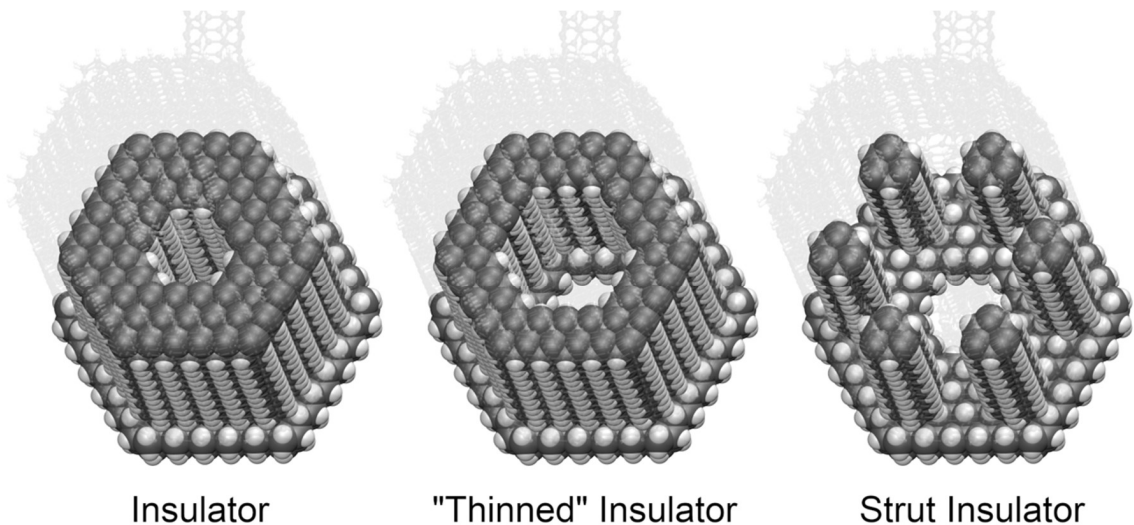

Insulator          "Thinned" Insulator          Strut Insulator

*Figure 13. Modifications that can be made to the proposed insulator design to approach more "vacuum-like" behavior by simple modification to the Lonsdaleite framework.*

To the issue of choice in insulator design, the DFT and DFTB calculations performed on the model channel-insulator structure (Figure 7) did reveal what might be a potential failure mode for the mFET in the form of electron tunneling from the insulator to the channel. HOMO and LUMO energies for the two components and for the full model structure are provided in Table 4, indicating a 0.54 eV (CAM-B3LYP) and 0.49 eV (B3LYP) difference in the HOMO energies for the insulator and channel, with the channel containing the lower-lying HOMO. By energy alone, the risk is that a less tightly bound electron from the insulator is predicted to preferentially transfer into the channel once a hole in the channel is present.

Table 4. HOMO and LUMO energies (in eV) for the model channel-insulator structure shown in Figure 7 and the difference in orbital energies for the two components ("Ch. – Ins. En$_{orbitals}$"). "B3LYP" = B3LYP/6-311G(2d,p)//B3LYP/3-21G*; "CAM-B3LYP" = CAM-B3LYP/6-311G(2d,p)//B3LYP/3-21G*.

|  | B3LYP | | CAM-B3LYP | |
|---:|:---:|:---:|:---:|:---:|
|  | HOMO | LUMO | HOMO | LUMO |
| *Full* | -5.4580 | 0.3859 | -6.8759 | 1.4422 |
| *Channel* | -5.9451 | 0.7339 | -7.4052 | 1.8667 |
| *Insulator* | -5.4523 | 0.6340 | -6.8697 | 1.7246 |
| *Ch. – Ins. En$_{orbitals}$* | -0.4928 | 0.0999 | -0.5355 | 0.1421 |

As an initial assessment of the likelihood of electron tunneling in the larger mFET, we can estimate the predicted orbital overlap of the channel and insulator combination in the model system by way of calculation of the electronic coupling integrals between the channel/insulator pair. Such an addition to the analysis workflow means that designs can be iterated on to reduce the likelihood of electron tunneling in some future mFET design. This can be done efficiently using the dimer projection method (DIPRO [23]), which estimates the electronic coupling between components of a system by projecting their wavefunctions onto dimer states. This approach has recently been incorporated into the semiempirical tight-binding code XTB [24,25], used here for the study of our model system with the GFN2-xTB method [26]. For the model system of a stacked co-facial benzene-nitrobenzene pair (Figure 14), the total orbital coupling obtained from the average of all individual pairs of orbital couplings, or $J_{AB,eff}$, is calculated to be 0.119 eV (hole transport in occupied molecular orbitals. See [25] for information about the accuracy of this approach against high-level DFT benchmarks). This same value for the small model channel/insulator system is 0.006 eV. For the large model system produced to test the increased physical separation of the channel and insulator components, this value is calculated to be 0.000 eV. As a first step in any more significant assessment of design pathologies related to the risk of electron tunneling in a design, this first set of calculations indicate that orbital overlap between the model channel/insulator combination might not lead to adverse behavior in a larger mFET design. That said, operational failures due to dynamic electron processes are among the most difficult and computationally demanding phenomena to calculate – the above assessment is merely a "first step" in what would necessarily be a much longer assessment and optimization process.

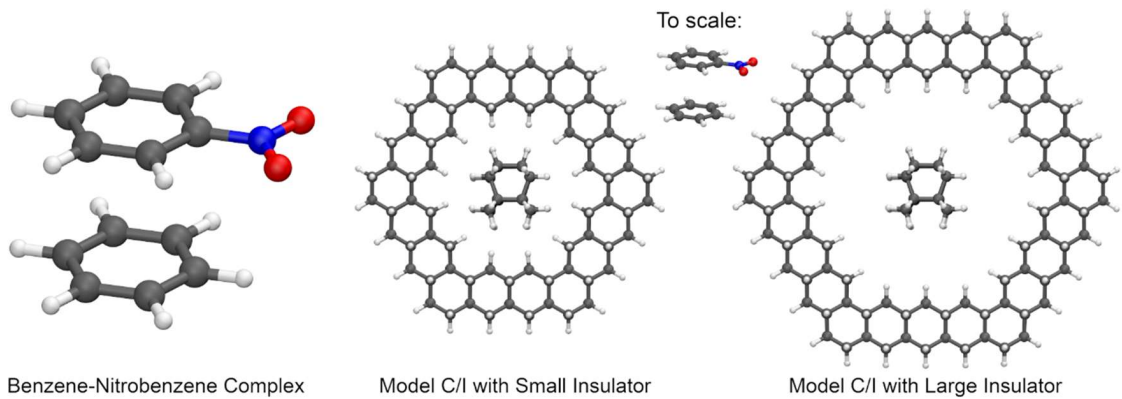

*Figure 14. Assessment of orbital overlap between the channel (C) and insulator (I) in our model test system, employing the DIPRO method within XTB to obtain $J_{AB,eff}$ for two variants of the C/I model and a benzene-nitrobenzene complex (shown enlarged and to scale with the C/I models) [27].*

An ever more radical approach would eliminate the insulator entirely by supporting the channel and the gate externally, so that there would be no material between them at all. For instance, based on an (18,18) CNT gate and the three-fold symmetry of the Lonsdaleite lattice, the possibility exists to link hexagonal Lonsdaleite blocks to either side of that (18,18) CNT gate and remove the insulator material completely. This would reduce the dielectric constant to 1 and the parasitic capacitance by a factor of 5.7. However, the impact of this on the breakdown voltage would have to be evaluated. The usual mechanism invoked for the breakdown voltage between two plates in vacuum involves acceleration of material between the two plates – but this mechanism assumes some first bit of material breaks off and is accelerated, leading to a cascade effect. If there is no first piece of material because the surfaces are small and atomically precise, then the mechanism for "breakdown" becomes less clear. Other than the obvious mechanism of tunneling between the gate and the channel and eventual thermal heating secondary to the tunneling current, it is not immediately obvious how "breakdown" could occur with such an atomically precise design.

## Resistance and Resistor-Capacitor (RC) Time Constant

Nanotubes have little intrinsic resistance along linear spans, however at junctions they typically have resistance in the vicinity of $R_k = 25,813 = h/e^2$, where $h$ is Planck's constant and $e$ is the charge of an electron. If we assume a ring oscillator composed of our mFETs uses about four junctions per FET, then we have R~100KΩ. With C ~ $10^{-18}$ F, this gives an RC time constant of ~$10^{-13}$ seconds, or 100 femtoseconds. This corresponds to a frequency of 10,000 gigahertz, or 10 terahertz. This is far infrared radiation with a wavelength in the range of 30 microns.

Such a FET could transmit and receive information at extremely high frequencies. Small numbers of such FETs could perform critical high-speed tasks, such as handling the extremely high frequency front-end processing of signals before handing off processing of the lower frequency result to more conventional devices. These and other applications are found in down-converters, multiplexers, demultiplexers, high speed analog-to-digital converters, very high-speed digital signal processing, and a host of other applications.

## Heat

If we assume that 1 V on the gate of this mFET is sufficient to switch it off, then because $Q = V\, C_g$ (where Q is the charge, V is the voltage and $C_g$ is the capacitance of the gate), we have $Q = V\, C_g = 1\, C_g = C_g = 10^{-18}$ C. The charge on a single electron is ~$1.6 \times 10^{-19}$ C. This means a charge of $10^{-18}$ C is ~ 6 electrons. Assuming we are switching near $10^{13}$ Hz (near the limiting switching speed as computed from the RC time constant), this implies a current flow of $10^{-18}$ C / $10^{-13}$ s, or $10^{-5}$ A. As heat generated is $I^2\, R$ and R is $10^5$ Ω, we have heat generated in continuous operation equaling $10^{-5}$ W, or 10 μW. Energy dissipated per switching operation is $10^{-5}$ W x $10^{-13}$ s, or $10^{-18}$ J.

To restate: a single mFET operating at one volt and switching as fast as possible generates ten microwatts of heat in continuous operation, or $10^{-18}$ J per switching operation.

Having defined key mFET characteristics, we can speculate about its possible impact on the performance of future computer systems.

As the volume of one mFET is 46.52 nm$^3$, we might generously assume that each mFET and associated wiring might occupy 1,000 nm$^3$ in a conceptual "ultimate mFET computer." Packed densely and operated at full speed, one cubic centimeter of this "ultimate mFET computer" would dissipate $(10^6)^3$ x $10^{-5}$ W, or $10^{13}$ W. The Hiroshima bomb, Little Boy, generated 63 TJ, or 6.3 x $10^{13}$ joules of energy. This hypothetical sugar-cube-sized computer switching at 10 THz would generate the energy of the Hiroshima blast every six seconds. Cooling would be challenging.

We conclude that many densely packed mFETs operating anywhere near their maximum speed will generate too much heat to remain operational for more than a short period of time because cooling will be either infeasible, too expensive, or both.

Heat is proportional to the square of the current. Switching speed is proportional to the current. If we halve the current and double the number of transistors, (1) each transistor will operate at half the speed while (2) the total number of switching operations per second will remain the same and (3) the total power used will be cut in half. Alternatively, we could halve the current and leave the number of transistors the same, in which case (1) each transistor will operate at half the speed and (2) the total number of switching operations per second would be cut in half and (3) the total power used would be cut by three fourths (75%). Pushing this further, we could operate our cubic centimeter computer $10^6$ times slower than 10 THz at 10 MHz, or 100 ns ($10^{-7}$ s) per switching operation. It would dissipate $10^{12}$ times less power than $10^{13}$ W, or 10 W. This gives us "only" $10^{18}$ x $10^7$ = $10^{25}$ switching operations per second, which is more than respectable for a 10 W home computer. This is $10^{-24}$ J per switching operation, or $10^{-17}$ W per transistor in continuous operation.

Although this energy-efficient mode involves slower switching (~100 ns, far slower than today's transistors), the vastly larger number of mFET devices operating in parallel ($10^{18}$, or a billion billion) would more than compensate. This hypothetical sugar-cube-sized computer would contain millions of times as many transistors as a powerful desktop computer today.

Achieving $10^{25}$ switching operations per second with a total system power of 10 W implies each switching operation dissipates only $10^{-24}$ J. This requires addressing several issues that have been well studied in the field of reversible computing, as $10^{-24}$ J << thermal noise at room temperature = $k_BT$, where $k_B$ is Boltzmann's constant and T is the temperature in Kelvins. At room temperature, $k_BT$ (T = 300 K) is ~ 4.14 x $10^{-21}$ J. This switching energy per operation is below the Landauer limit, meaning that a conventional (irreversible) logic device would be thermodynamically impossible to operate at this power level. This highlights the need to use reversible computing to achieve such densities. [28]

It should be noted that these numbers are rough approximations. It should also be noted that this mFET is a preliminary design and unlikely to be optimal. Further improvements and refinements, some noted here and others obvious to those skilled in the art, would significantly improve its performance. With the advent of semiempirical methods such as DFTB that enable the analysis of multi-thousand atom systems on commodity hardware with near-DFT accuracy for many calculable properties, a rigorous assessment and optimization cycle based on simulation and experimental feedback is believed to be well within near-term engineering capabilities for aspects of the proposed design.

## Conclusion

We have analyzed an atomically precise P-channel molecular FET (mFET) with 7,694 atoms made only of hydrogen and carbon atoms. It uses metallic (4,4) carbon nanotubes as the conductive leads, a linear segment of Lonsdaleite (hexagonal diamond) as the channel, Lonsdaleite as the insulating layer between the channel and the gate, and a (20,20) metallic carbon nanotube as the surrounding gate. Its volume is ~46 nm$^3$.

Its predicted peak switching speed is ~100 femtoseconds. When switching at full speed, it would then use ~$10^{-18}$ J per switching operation. When switching speed is slowed to ~100 ns, it generates only ~$10^{-24}$ J per switching operation when used in a properly designed circuit. A sugar-cube-sized computer with $10^{18}$ mFETs could dissipate as little as 10 watts while delivering ~$10^{25}$ switching operations per second.

# References

1.  Halliday, D. and Resnick, R., Fundamentals of Physics (Extended 3rd Edition), Wiley, 1988.
2.  Sze, S. M., Physics of Semiconductor Devices, Second Edition, Wiley-Interscience, 1981.
3.  Sze, S. M., editor, High Speed Semiconductor Devices, Wiley-Interscience, 1990.
4.  Z. Chen, D. Farmer, S. Xu, R. Gordon, P. Avouris, and J. Appenzeller, "Externally assembled gate-all-around carbon nanotube field-effect transistor," *IEEE Electronic Device Letters*, p. 183-5, vol. 29(2) (2008), docs.lib.purdue.edu/nanodocs/170.
5.  J. Appenzeller, "Carbon Nanotubes for High-Performance Electronics - Progress and Prospect," *Proceedings of the IEEE*, p. 201-211, vol. 96(2) (2008), 10.1109/JPROC.2007.911051.
6.  R. Nizam and M. M. Sehban, "Calculating the Band Gaps of Perfect Carbon Nanotube through Tight Binding Method." *International Journal of Scientific Research*, p. 1565-70, vol. 6(1) (2017), www.ijsr.net/archive/v6i1/ART20164401.pdf.
7.  G. Gelao, R. Marani, and A. G. Perri, "A Formula to Determine Energy Band Gap in Semiconducting Carbon Nanotubes," *ECS Journal of Solid State Science and Technology*, p. M19-M21, vol. 8(2) (2019), 10.1149/2.0201902jss.
8.  Gaussian 16, Revision C.01, Gaussian, Inc., Wallingford CT, 2013, www.gaussian.com.
9.  A. D. Becke, "Density-functional thermochemistry. III. The role of exact exchange," *J. Chem. Phys.*, p. 5648-52, vol. 98 (1993), 10.1063/1.464913.
10. Hariharan, P. C., Pople, J. A. "The influence of polarization functions on molecular orbital hydrogenation energies." *Theor. Chim. Acta*, p. 213-222, vol. 28 (1973), 10.1007/bf00533485.
11. Hehre, W. J., Ditchfield, R., Pople, J. A. Self-Consistent Molecular Orbital Methods. XII. Further Extensions of Gaussian-Type Basis Sets for Use in Molecular Orbital Studies of Organic Molecules." *J. Chem. Phys.*, p. 2257-2261, vol. 56, 10.1063/1.1677527.
12. Krishnan, R., Binkley, J. S., Seeger, R., Pople, J. A. "Self-consistent molecular orbital methods. XX. A basis set for correlated wave functions." *J. Chem. Phys.* p. 650-654, vol. 72 (1980), 10.1063/1.438955.
13. W. S. Su, T. C. Leung, and C. T. Chan, "Work function of single-walled and multiwalled carbon nanotubes: First-principles study," *Phys. Rev. B*, p. 235413, vol. 76 (2007), 10.1103/PhysRevB.76.235413. *"The calculations are performed within the local density approximation (LDA) framework using the Ceperley-Alder form of exchange-correlation functional and ultrasoft pseudopotentials with a plane-wave cutoff of 358 eV."*
14. CVD Diamond - Properties, www.cvd-diamond.com/properties_en.htm.
15. Technologies and Applications — International FemtoScience, Inc. (FemtoSci), www.femtosci.com/applications.
16. M. Elstner, D. Porezag, G. Jungnickel, J. Elsner, M. Haugk, T. Frauenheim, S. Suhai, and G. Seifert, "Self-consistent-charge density-functional tight-binding method for simulations of complex materials properties," *Phys. Rev. B*, p. 7260, vol. 58 (1998), 10.1103/PhysRevB.58.7260.
17. B. Hourahine, B. Aradi, V. Blum, F. Bonafé, A. Buccheri, C. Camacho, C. Cevallos, M. Y. Deshaye, T. Dumitrica, A. Dominguez, S. Ehlert, M. Elstner, T. van der Heide, J. Hermann, S. Irle, J. J. Kranz, C. Köhler, T. Kowalczyk, T. Kubar, I. S. Lee, V. Lutsker, R. J. Maurer, S. K. Min, I. Mitchell, C. Negre, T. A. Niehaus, A. M. N. Niklasson, A. J. Page, A. Pecchia, G. Penazzi, M. P. Persson, J. ˇRezác, C. G. Sánchez, M. Sternberg, M. Stöhr,

F. Stuckenberg, A. Tkatchenko, V. W.-z. Yu, and T. Frauenheim. "DFTB+, a software package for efficient approximate density functional theory based atomistic simulations." *J. Chem. Phys.*, p. 124101, vol. 152 (2020), 10.1063/1.5143190.
18. E. Rauls, R. Gutierrez, J. Elsner, and Th. Frauenheim, "Stoichiometric and Non-Stoichiometric (10-10) and (11-20) surfaces in{2H-SiC}: a theoretical study," *Sol. State Comm.*, p. 459-64, vol. 111(8) (1999), 10.1016/S0038-1098(99)00137-4.
19. E. Caldeweyher, C. Bannwarth, and S. Grimme, "Extension of the D3 dispersion coefficient model," *J. Chem. Phys.* p. 034112, vol. 147 (2017), 10.1063/1.4993215.
20. S. Grimme, S. Ehrlich and L. Goerigk, "Effect of the damping function in dispersion corrected density functional theory," *J. Comp. Chem.*, p. 1456-65, vol. 32(7) (2011), 10.1002/jcc.21759.
21. T. Yanai, D. Tew, and N. Handy, "A new hybrid exchange-correlation functional using the Coulomb-attenuating method (CAM-B3LYP)," *Chem. Phys. Lett.*, p. 51-57, vol. 393(1-3) (2004), 10.1016/j.cplett.2004.06.011.
22. M. Sims, B. Smith, B. Helfrich, N. Sathaye, E. Messick, T. Moore, R. Fish, M. Rajagopalan, P. Rotkiewicz, D. Hendricks, K.E. Drexler, and D. G. Allis, "NanoEngineer-1 - A CAD-based molecular modeling program for structural DNA nanotechnology," *Foundations of Nanoscience 2008 (FNANO08)*, Snowbird, UT, USA, April 22-25, 2008. Available from github.com/kanzure/nanoengineer.
23. B. Baumeier, J. Kirkpatrick, and D. Andrienko, "Density-functional based determination of intermolecular charge transfer properties for large-scale morphologies," *Phys. Chem. Chem. Phys.*, p. 11103-13, vol. 12 (2010), doi.org/10.1039/C002337J.
24. C. Bannwarth, E. Caldeweyher, S. Ehlert, A. Hansen, P. Pracht, J. Seibert, S. Spicher, and S. Grimme, "Extended tight-binding quantum chemistry methods," *WIREs Comput. Mol. Sci.*, p. e1493, vol. 11(2) (2020), 10.1002/wcms.1493.
25. J. T. Kohn, N. Gildemeister, S. Grimme, D. Fazzi, and A. Hansen, "Efficient calculation of electronic coupling integrals with the dimer projection method via a density matrix tight-binding potential," *J. Chem. Phys.*, p. 144106, vol. 159 (2023), doi.org/10.1063/5.0167484.
26. C. Bannwarth, S. Ehlert, and S. Grimme, "GFN2-xTB—An Accurate and Broadly Parametrized Self-Consistent Tight-Binding Quantum Chemical Method with Multipole Electrostatics and Density-Dependent Dispersion Contributions," *J. Chem. Theory Comput.*, p. 1652-71, vol. 15(3) (2019), 10.1021/acs.jctc.8b01176.
27. For example, discussion, and code, see xtb-docs.readthedocs.io/en/latest/dipro.html.
28. M. P. Frank, "Throwing computing into reverse," *IEEE Spectrum*, p. 32-37, vol. 54(9) (2017), doi:10.1109/MSPEC.2017.8012237.